\documentclass[a4paper,12pt]{article}

\begin{document}
\vspace{5 pt}

\centerline{\large \bf On the physical nature of the fluorescence depletion 
effect}

\vspace{7 pt}
\centerline{\sl V.A.Kuz'menko}
\vspace{5 pt}
\centerline{\small \it Troitsk Institute for Innovation and Fusion Research,}
\centerline{\small \it Troitsk, Moscow region, 142190, Russian Federation.}
\centerline{\small \ E-mail: kuzmenko@triniti.ru}

\vspace{5 pt}
\begin{abstract}

The physical nature of the fluorescence depletion (FD) effect and 
the possible way of  its experimental study are discussed.

\vspace{5 pt}
\end{abstract}

\vspace{12 pt}

The FD effect is used for many years as an efficient time-domain 
method of rotational spectroscopy for large polyatomic molecules [1-3]. 
In this method, the picosecond laser pulse is split into the pump and the 
probe pulses, which are focused at the molecular jet. Time and frequency-
integrated fluorescence intensity from the sample is studied as a function 
of the delay between the pump and the probe laser pulses. The observed 
rotational recurrences allow to obtain the corresponding rotational 
constants of the ground and excited states of the molecule [4-6].

However, the physical nature of the FD effect is practically 
unknown for present day. There exists rather formal phenomenological 
description, which is based on the concepts of coherency, quantum 
interference and quantum beats [1]. It is worth to note, that {\bf this 
situation is rather typical in nonlinear optics, where phenomenological 
description often takes the place of the physical explanation}. As a 
result, some problems appear, if one wants to understand the physical sense 
of such a descriptions [7].

A real physical explanation of the FD effect must be based on the 
supposition that the laser pulse ``interacts preferentially with those species 
whose transition dipoles are aligned along its polarization vector at the 
instant that it passes through the sample" [1]. But this correct idea should 
be extended further: the absorption cross-section of laser radiation by 
molecules depends on the orientation of the molecule in space, on the 
angle between the axes of the molecule and that of the laser beam. How 
strong is this dependence? How sharp is it? These are obviously the most 
interesting questions now. The FD effect provides us an excellent possibility 
to obtain such information. It is surprising that no one has studied the 
nature of FD effect till now.

In all cases extremely intense pump and probe laser pulses of equal 
energy were used [8]. This situation meets the needs of the four photon 
mixing effect, but it isn't suited to the study of the physical nature of the 
effect. The intensity of the pump beam should be usually smaller, than the 
saturation intensity. The intensity of the probe beam should be still 
substantially smaller (0,01--- 0,1 of the pump beam). It shouldn't  perturb 
the system. In these conditions an efficient lock-in technique can be used 
[9,10]. Besides the usual pump-probe delay dependence of the 
fluorescence intensity, the main goal of such experiments should be the 
study of the dependence of the FD effect on the angle between the pump 
and the probe beams [8]. For high intensity regime, the latter dependence 
is expected to be relatively flat. But for the low intensity regime, when 
four-photon mixing effects can be suppressed, the dependence of the FD 
effect on the angle between the two laser beams is expected to be sharper.

Careful experimental study of the FD effect may be also very 
useful for studying the physical nature of the so-called population transfer 
effect with counterintuitive sequence of interactions with two radiation 
pulses (CSI) [11,12]. In this case, the most efficient population transfer 
between two low lying states takes place when the Stokes pulse overlaps 
and precedes the pump one. Again, we have a phenomenological 
description of this effect [13,14], but its physical explanation is absent till 
now. The CSI effect takes place even when broad femtosecond laser 
pulses interact with large polyatomic molecules in a condensed phase 
[7,15]. In this case the CSI effect is the first step of the four photon 
mixing process, and the laser pulses duration determines the value of the 
so-called peak shift [15-18].

Thus, experimental investigation of the physical nature of the FD 
effect is much more important and interesting, than simple study of the 
rotational spectra of large polyatomic molecules.

\end{document}